\documentclass[12pt]{article}
\begin{document}
\title
{Dielectric crystal in the Planck blackbody}
\author{Miroslav  Pardy\\[0.2cm]
{\it Department of Physical Electronics}, \\
Masaryk University, Faculty of Science, Kotl\'{a}\v{r}sk\'{a} 2,\\ 611 37 Brno, Czech Republic \\
{\it e-mail: pamir@physics.muni.cz}\\
\date{\today}}
\maketitle

\vskip 5mm
\large
\begin {abstract}
The dielectric crystal with the index of refraction $n$ is inserted in the Planck blackbody. The spectral formula for photons in such dielectric medium is derived with the equation for the temperature of internal photons. The derived equation is solved for the constant index of refraction. The photon flow initiates the osmotic pressure of he Debye phonons in the dielectric blackbody.  
\end{abstract}

\vskip 3mm

\baselineskip 17 pt

\section{Introduction}

It is physically meaningful to consider, in quantum theory of light and quantum theory of solids, dielectric crystalline medium with phonons which is  inserted in the Planck blackbody photon gas. It means that photon gas of the blackbody surrounding  the dielectric crystalline medium with with index of refraction $n$ flows into such crystal and initiate the quantum osmotic pressure of  photons as solvent and phonons as solute.

The classical osmosis is the spontaneous passage  of solvent molecules through a partially
permeable membrane separating two solutions of different concentration into a region of higher
solute concentration of solute, in order to equalize the solute concentrations on 
the two sides. The physical law which controll the osmotic pressure is
so called van't Hoff's equation (published in 1885):

$$p = i \frac{C}{\mu}RT,\eqno(1)$$
where $p, i, C, \mu, R, T$ are  pressure, van't Hoff factor, concentration of solute, mollar mass, thermodynamic gas constant and temperature, and concentration is defined by formula $C = m/V$, where $m$ is mass of solute in volume $V$. We consideer here the quantum osmosis with photons and phonons and wizh the semi-permeable membrane for photons which is the surface of the dielectric crystal. 

The derivation of the van't Hoff formula using the thermodynamic potential can be found in the texbooks on thermodynamics and statistical physics 
(Landau et al., 1980). The derivation of the osmotical pressure from rigorous statistical physics was given by Isihara (1971). On the other hand, the quantum theory of osmosis was not published. A Duth physical and organic chemist van't Hoff presented his Nobelian theory long time before the introduction of photons into physics by Max Planck, Lewis and Einstein and before the introduction of phonons into solid state physics by Einstein and Debye. So, the problem of the osmotic pressure in the Planck blackbody with the dielectric medium arises as the problem of modern physics.  

The dielectric crystal  with photons is called here  by term Planck dielectric blackbody. Inside of the dielectric medium with index of refraction $n$, the spectral radiation formula is modified and we derive  in the next part mathematical form of the spectrum of such dielectric blackbody. The derivation of the spectral formula is based on the original Planck spectral formula which was rederived by Einstein (1917). 

\section{The Einstein blackbody model}

The distribution of the blackbody photons was derived by Planck (1900) from modification of the 
thermodynamical entropy, and later,  Einstein (1919) derived the Planck formula from the Bohr model of
 atom which was based on two postulates: 1. every atom can exist in the discrete series of states in
 which electrons do not radiate even if they are moving at acceleration  (the postulate of the
 stationary states), 2. transiting electron from the stationary state to other, emits the energy
 according to the law $\hbar\omega = E_{m} - E_{n}$, called the Bohr formula, where $E_{m}$ is 
the energy of an electron in the initial state, and $E_{n}$ is the energy of the final state of
 an electron to which the transition is made and $E_{m} > E_{n}$.

Let us remark still that the Bohr theory does not involve the physical mechanism of creation of photons
 and the adequate model of photon. However, it follows from  quantum theory of fields,  that photon
 is excited state of vacuum and at the same time also an electron is the excited state of vacuum, 
which follows from the elementary experimental equation
 $\gamma + \gamma \rightleftharpoons e^{+} + e^{-}$ (Berestetzkii et al., 1999).  
At present time we know from the most general quantum field theory that all matter
 and antimatter in universe are excited states of vacuum.

Einstein introduced coefficients of spontaneous  and stimulated emission
$A_{mn}, B_{mn}, B_{nm}$. In case of spontaneous emission,
the excited atomic state decays without external stimulus as an analogue
of the natural radioactivity decay. The energy of the emitted photon
is given by the Bohr formula. In the process of the stimulated
emission the atom is induced by the external stimulus to make the
same transition. The external stimulus is a blackbody photon that has
an energy given by the Bohr formula.

If the number of the excited atoms is equal to $N_{m}$, the emission
energy per unit time  conditioned by the spontaneous transition from
energy level ${E_{m}}$ to energy level ${E_{m}}$ is

$$P_{spont. \; emiss.} = N_{m}A_{mn} \hbar\omega, \eqno(2) $$
where $A_{mn}$ is the coefficient of the spontaneous emission.

In case of the stimulated emission, the coefficient  $B_{mn}$
 corresponds to the transition of an electron
 from energy level ${E_{m}}$ to energy level ${E_{n}}$ and coefficient
 $B_{nm}$ corresponds to the transition of an electron
 from energy level ${E_{n}}$ to energy level ${E_{m}}$. So, for the
 energy of the stimulated emission per unit time we have two formulas :

$$P_{stimul. \; emiss.} = \varrho_{\omega}N_{m}B_{mn} \hbar\omega \eqno(3) $$

$$P_{stimul. \; absorption} = \varrho_{\omega}N_{n}B_{nm}
\hbar\omega. \eqno(4) $$

If the blackbody is in thermal equilibrium, then the number of
transitions from ${E_{m}}$ to ${E_{n}}$ is the same as 
from ${E_{n}}$ to ${E_{m}}$ and we write:

$$N_{m}A_{mn}\hbar\omega + N_{m}\varrho_{\omega}B_{mn}\hbar\omega =
N_{n}\varrho_{\omega}B_{nm}\hbar\omega,  \eqno(5)$$
where $\varrho_{\omega}$ is the density of the photon energy of the blackbody. 

Then, using the Maxwell statistics

$$N_{n} = De^{-\frac{E_{n}}{kT}},\quad  N_{m} = De^{-\frac{E_{m}}{kT}},
\eqno(6)$$
we get:

$$\varrho_{\omega} = \frac{\frac{A_{mn}}{B_{mn}}}
{\frac{B_{nm}}{B_{mn}}e^{\frac{\hbar\omega}{kT}} - 1}.\eqno(7)$$

The spectral distribution of the blackbody does not depend on the
specific atomic composition of the blackbody and it means the formula
(7) must be so called the Planck formula:

$$\varrho_{\omega} = \frac{\hbar\omega^3}{\pi^2 c^3}\frac{1}
{e^{\frac{\hbar\omega}{kT}} - 1}.\eqno(8)$$

After comparison of eq. (7) with eq. (8) we get:

$$B_{mn} = B_{nm} = \frac{\pi^2 c^3}{\hbar \omega^3}A_{mn}. \eqno(9)$$

It means that the probabilities
of the stimulated transitions from ${E_{m}}$ to ${E_{n}}$ 
and from ${E_{n}}$ to ${E_{m}}$ are
proportional to the probability of the spontaneous transition
$A_{mn}$. So, it is sufficient to determine only one of the
coefficient in the description of the radiation of atoms.

The internal density energy  of the blackbody gas is given by integration of the last equation over all  frequencies $\omega$, or

$$u = \int_{0}^{\infty}\varrho(\omega)d\omega  = a T^{4}; \quad a = \frac{\pi^{2}k^{4}}{15\hbar^{3}c^{3}}\eqno(10)$$
and the pressure of photons inside the blackbody follows from the electrodymanic situation inside blackbody as follows:

$$p = \frac{u}{3}\eqno(11).$$

Let us remark that coefficients $A_{mn}$ of the so called spontaneous
emission cannot be specified in the framework of the classical 
thermodynamics, or, statistical physics. They  can be determined 
only by the methods of quantum
electrodynamics as the consequences of the so called radiative
corrections. So, the radiative corrections are hidden external stimulus, 
which explains the spontaneous emission.

 \section{The dielectric blackbody}

We suppose here that inside of the Planck blackbody there is the dielectric crystal with the index of refraction $n(\omega)$. Then, the wave vector of photon inside the dielectric medium  is given by known formula

$$q = n(\omega)\frac{\omega}{c}. \eqno(12)$$

The number of light modes in the interval $q, q + dq$ inside of the dielectric in the volume $V$
is  $Vq^{2}dq/\pi^{2}$. After differentiation of formula (12) we get with 
$d\ln\omega = d\omega/\omega $

$$dq = \frac{1}{c}[n(\omega) + \omega \frac{dn(\omega)}{d\omega}]d\omega = 
\frac{n(\omega)}{c}\frac{d\ln [n(\omega)\omega]}{d\ln\omega}d\omega.\eqno(13)$$

Then, it is easy to see that the number of states  in the interval $\omega, \omega + d\omega$ 
of the electromagnetic vibrations in the volume $V$ is 

$$Vg(\omega)d\omega = \frac{V}{\pi^{2}}\left(\frac{n(\omega)}{c}\right)^{3}\frac{d\ln [n(\omega)\omega]}{d\ln\omega} d\omega. \eqno(14)$$

If we multiply the last formula by the average energy of the harmonic oscillator,

$$<E_{\omega}>  = \frac{\hbar\omega} {e^{\frac{\hbar\omega}{kT}} - 1},
\eqno(15)  $$
we get the Planck formula for the blackbody with dielectric medium:

$$\varrho(\omega) = \frac{n^{3}(\omega)\omega^{2}}{\pi^{2}c^{3}}
\frac{d\ln [n(\omega)\omega]}{d\ln\omega}
\frac{\hbar\omega} {e^{\frac{\hbar\omega}{kT}} - 1},
\eqno(16)$$
where for $n = 1$, we get exactly formula (8).

\section{The oscillator model of the index of refraction}  
                 
This model follows from the classical theory of dispersion, which is based on the vibration equation of electron in an atom  

$$\ddot x  + \gamma\dot x + \omega_{0}^{2}x = \frac{e}{m}E_{0}\cos\omega t,\eqno(17)$$
where $\gamma$ is the oscillator constant and $\omega_{0}$ is the basic frequency of oscillator. The symbol  $\omega$ is the frequency of the applied electric field. The index of refraction  following from eq (17) is given by the formula (Garbuny, 1965)

$$n = 2\pi N\frac{e^{2}}{m}\frac{\omega_{0}^{2} - \omega ^{2}}{(\omega_{0}^{2}- \omega^{2})^{2} +\gamma^{2}\omega^{2}},\eqno(18)$$
where  $N$ is number of electrons in the unit of volume.

In case of electrons with basic frequencies $\omega_{1}, \omega_{2}, \omega_{3}, \omega_{4} ...  \omega_{n}$, the last refraction index can be generalized to form more complex mathematical object. We consider here, to be  pedagogical clear,  only one oscillator with one basic frequency. Nevertheless it is possible consider arbitrary dielectric material with the phenomenological index of refraction.

Now the question arises, if the dielectric blackbody can be considered as the solution composed from atoms, phonons and photons where the osmotic pressure play some role. 
We had accepted this hypothesis as the correct one.

\section{The osmosis in dielectric blackbody}

Phonons were introduced in the crystal physics by Einstein in order to derive the adequate formula for he specific heat. The Einstein formula was generalized and improved by Debye
who derived the formula for the average energy of phonons in a crystal in the interval of temperatures $\Theta-\delta < T < \Theta + \delta$ ($\delta$ is some parameter) as follows (Rumer et al., 1977):

$$ U = N\varepsilon_{0} + 3NTD\left(\frac{\Theta}{T}\right),\eqno(19)$$ 
where $\varepsilon_{0} = (9/8)\hbar \omega_{max}$, where 

$$\omega_{max} = 2\pi v\left(\frac{3N}{4\pi V}\right)^{1/3}\eqno(20)$$
and $D(x)$ is so called the Debye wave function of the following structure:

$$D(x) = \frac{3}{x^{3}}\int_{0}^{x}\frac{y^{3}}{e^{y} - 1}dy,\eqno(21)$$
and the critical temperature $\Theta$ was derived by Debye in the following form:

$$\Theta = v\left(\frac{6\pi^{2}N}{V}\right)^{1/3},\eqno(22)$$
with $v$ being velocity of sound waves defined in the theory of elasticity of the crystal. 
 
Let us compare the internal energies of the pure blackbody and dielectric blackbody and then let us compare the pressure inside of the pure blackbody and inside the dielectric blackbody.

For pure blackbody, we have $u = aT^{4}$ and for model with 
$n$ given by eq. (18) we have 

$$u = \int_{0}^{\infty}\varrho_{n}(\omega)d\omega =    \int_{0}^{\infty}\varrho_{n}(\omega)\frac{n^{3}(\omega)\omega^{2}}{c^{3}}\frac{d\ln [n(\omega)\omega]}{d\ln \omega}
\frac{\hbar\omega} {e^{\frac{\hbar\omega}{kT}} - 1}d\omega.\eqno(23)$$

Because the dielectric medium is permeable for photons and not for phonons (the photon osmosis), the outer pressure is equal to the photon gas pressure in the dielectric blackbody, or  $p(n) = u(n)/3  = u/3$. So,

$$ \int_{0}^{\infty}\varrho_{n}(\omega)d\omega =  u/3 = \frac{aT^{4}}{3}, \eqno(24)$$ 
or, 

$$\int_{0}^{\infty}\frac{n^{3}(\omega)\omega^{2}}{\pi^{2}c^{3}}\frac{d\ln [n(\omega)\omega]}{d\ln\omega} \frac{\hbar\omega} {e^{\frac{\hbar\omega}{kT_{diel}}} - 1}d\omega =
 \frac{aT^{4}}{3},\eqno(25)$$
where we introduced the dielectric  temperature $T_{diel}$, which physically means that the temperature of dielectric blackboddy is not the same as the temperature of the bath of vacuum blackbody photons. The last equation is the integral equation for function $T_{diel}$ and in general represents very difficult mathematical problem of the future physics of the dielectric blackbody. The experimental verification of the last equation will be also the crutial problem of photon phyics. 

In the most simple case with $n = const$,
we get fter some algebraic operation, that the temperature dielectric blackbody
surraunded by the vacuum blackbody  is given by the formula 

$$T_{diel} = \frac{T}{\sqrt[4]{n^{3}}}.\eqno(26)$$

The last formula can form the goal of the experimenters working in the blackbody radiation physics.
The dielectric as the osmotic membrane plays the role of the Maxwell demonic refrigerator. The second possibility is to put $n = n(T)$ in order to get the integral equation for the dependence of the index of refraction on temperature. However, it seems that this assumption is not physically adequate.

In case of the dielectric Debye crystal, the equation of state is (Rumer et al., 1977)

$$p = \left(\frac{U_{phon}}{\Theta} - \frac{9}{4}N\right)\frac{d\Theta}{dV},\eqno(27)$$
where $V$ and $N$ is volume and number of oscillators in crystal. 
The difference $\Delta p =  p(T) - p(T_{diel})$ is the osmotic pressure caused by the photon flow. 

In case of the two-dimensional crystal, the internal phonon energy is (Rumer et al., 1977)

$$U_{2D-phon} = \frac{4}{3}N \Theta \left[1 + \left(\frac{T}{\Theta}\right)^{3}
\int_{0}^{\Theta/T}\frac{y^{2}}{e^{y} - 1}dy\right].\eqno(28)$$
and 

$$\Theta = 2\pi v\left(\frac{N}{\pi \sigma}\right)^{1/2},\eqno(29)$$ 
where $\sigma$ is  the area of the 2D crystal (e. g. graphene, which is the carbon sheet),
instead of $d\Theta/dV$ is $d\Theta/d\sigma$ and $9/4$ must be replaced by the adequate constant.
The osmotic temperature  of the 2-dimensional and 1-dimensional dielectric crystal is an analogue of the 3-dimensional case and can be derived from the formulas by author article (Pardy, 2013).

\section{Discussion}

The clssical osmosis is the physical phenomenon in the system with solute, solvent, solution and semi-permeable membrane. It plays fundamental role in biological and physiological systems, where for instance the photosynthesis in plants is not posible without water and photon osmosis and human being  does not exist without liquid osmosis.

Isihara (1971) derived from the statistical physics the following formula for the osmotic pressure of the two-component statistical system:

$$p = kT\frac{\partial[\ln(\Xi/\Xi_{0})]}{\partial V},
\eqno(30)$$
where $\Xi$ and $\Xi_{0}$ are the big statistical sums of solute and solvent. The explicit mathematical form of the formula is sophisticated and the derivation of the van't Hoff formula is not elementary.

We have generalized the classical osmosis to the photon osmosis with phonons and photons 
where the osmotical pressure is realized by phonons and photons inside the medium with index of refraction.
The change of temperature caused by osmotical pressure was described by eq. (25), which was solved by us only for the most simple case of the constant index of refraction. The solution of the general case is the problem of the future osmotic and photonic physics for the arbitrary index of refraction. The dielectric surface is the osmotic semi-permeable membrane and plays the role of the Maxwell demonic refrigerator. 

The theory of phonon-photon dielectric blackbody is the preamble for experiments  for the determination of the osmotic process as the consequence of the quantum properties of the phonon-photon gas. The role of phonon-photon osmosis  in biological and physiological systems is crucial. The  phonon-photon osmotic pressure plays probably substantional negative role in the formation  and in the development of skin cancer. 
  
 It is not excluded, that the experiments with the quantum osmosis in plasma with magnetic field as semi-permeable osmotic membrane, will play crucial role in  the TOKAMAK 
fusion reactor  physics. 

\vspace{15mm}

References

\vspace{5mm}

\noindent
Berestetzkii, V. B.,  Lifshitz E. M. and Pitaevskii, L. P. \\  {\it Quantum electrodynamics},
(Butterworth-Heinemann, Oxford, 1999).\\[2mm]
Einstein, A. (1917). Zur quantentheorie der Strahlung, \\{\it Physikalische Zeitschrift}, {\bf 18}, 121.\\[2mm]
Isihara, A.  {\it Statistical Physics}, (Academic Press, New York-London, 1971). \\[2mm]
Kubo, R. {\it Statistical Mechanics}, (Norh-Holland Publishing Company - Amsterdam, 1965). \\[2mm]
Landau, L. D. and Pitaevskii, {\it Statistical Physics}, Volume 5, \\Third edition, revised
and enlarged, (Pergamon Press, Oxford, New York,.., 1980). \\[2mm] 
Pardy, M. (1913). Velocity of sound in the blackbody photon gas, {\it Results in physics} {\bf 3}, 7073.\\[2mm]
Planck, M. (1900). Zur Theorie des Gesetzes der Energieverteilung im
Normalspektrum, {\it Verhandlungen deutsch phys., Ges.}, {\bf 2}, 237.; \\
ibid: (1901). {\it Ann. Phys.}, {\bf 4}, 553. \\[2mm]
Rumer Yu. B.  and  Ryvkin, M. Sch. {\it Thermodynamics, statistical physics, \\kinetics}, (Nauka, Moscow, 1977), (in Russian).
\end{document}